\documentclass[default]{sn-jnl}

\usepackage{graphicx}%
\usepackage{multirow}%
\usepackage{amsmath,amssymb,amsfonts}%
\usepackage{amsthm}%
\usepackage{mathrsfs}%
\usepackage[title]{appendix}%
\usepackage{xcolor}%
\usepackage{textcomp}%
\usepackage{manyfoot}%
\usepackage{booktabs}%
\usepackage{algorithm}%
\usepackage{algorithmicx}%
\usepackage{algpseudocode}%
\usepackage{listings}%

\usepackage{mathtools}
\usepackage{lineno}

\theoremstyle{thmstyleone}%
\theoremstyle{thmstyletwo}%

\theoremstyle{thmstylethree}%

\begin{document}

\title{Application of the Most Frequent Value Method for $^{39}$Ar  Half-Life Determination}



\author*[1]{\fnm{Victor V.} \sur{Golovko}}\email{Victor.Golovko@cnl.ca}

\affil*[1]{
	\orgname{Canadian  Nuclear  Laboratories}, \orgaddress{\street{286 Plant Road}, \city{Chalk  River}, \postcode{K0J 1J0}, \state{Ontario}, \country{Canada}}}

\date{}

\abstract{
An evaluation method supported by robust statistical analysis was applied to  historical measurements of $^{39}$Ar half-life. The method, based on the most frequent value (MFV) approach combined with bootstrap analysis, provides a more robust way to estimate $^{39}$Ar half-life, and results in $T_{1/2}($MFV$) = 268.2^{+3.1}_{-2.9}$ years with uncertainty corresponding to the 68\% confidence level. The uncertainty is a factor of 3 smaller than that of the most precise re-calculated $^{39}$Ar half-life measurements by Stoenner et al., (1965) and  a factor of 2.7 smaller than that of the adopted half-life value in nuclear data sheets. Recently, the specific activity of the beta decay of $^{39}$Ar in atmospheric argon was measured in several underground facilities. Applying the MFV method to a specific activity of  $^{39}$Ar from underground measurements results in $ SA_{{^{39}\text{Ar}}/\text{Ar}}(\text{MFV}) = 0.966^{+0.010}_{-0.018} \, \, \text{Bq/kg$_{\text{atmAr} } $} $ with uncertainty corresponding to the 68\% confidence level.  In this paper the method to determine the half-life of $^{39}$Ar using the specific activity of $^{39}$Ar in atmospheric argon is also discussed.
}


\keywords{Most Frequent Value, Robust statistical analysis,  $^{39}$Ar half-life, Bootstrap analysis, Specific activity of $^{39}$Ar}

\maketitle



\section{Introduction}
\label{intro}

The half-life is one of the fundamental properties of radioactive nuclei. However, there are many questions for isotopes having a half-life of around a hundred years or more, for which the determination of the decay curve becomes difficult and the measurement is very long. Moreover, if the decay curve for a particular isotope is not observed for a few half-lives (typically 3--4 or more) one could deduce very strange results (see for example Spillane et al. (2007)~\cite{spillane2007198}, where the decay curve was observed for about one half-life). The results from Spillane et al., (2007)~\cite{spillane2007198} were not confirmed when the decay curve was measured for a few half-lives with a much better accuracy by Goodwin et al. (2007)~\cite{goodwin2007half}.

When a cosmic ray strikes atoms in the stratosphere one of several interactions may occur that produce radioargon. Here we are interested in the $^{39}$Ar isotope, which is produced mainly via the $^{40}$Ar(n, 2n) process \cite{saldanha2019cosmogenic,zhang2022evaluation}. When interaction products reach the Earth's surface another process may affect the rate of radioargon production, for example via the $^{39}$K(n, p)$^{39}$Ar process. However, based on the amount of stable argon isotopes in the atmosphere, the radioargon production from Earth's surface is significantly smaller. Moreover, noble gas nuclides produced in the Earth's surface follow transport processes in the environment. These geophysical and geochemical properties are used in various tracer applications in the geosciences \cite{loosliUse39Ar14C1980,LEHMANN1997727,luTracerApplicationsNoble2014}. These properties are especially crucial in  practical applications for $^{39}$Ar dating methods in the field of isotope geochronology~\cite{renne1998intercalibration}. The largest problem for the $^{39}$Ar dating method is the uncertainty of the half-life involved. As further discussion shows, with the approach used in this work the uncertainty on the half-life of $^{39}$Ar is reduced by factor of 3 compared to that of the most precise re-calculated $^{39}$Ar half-life measurements by Stoenner et al. (1965)~\cite{stoennerHalflivesArgon37Argon391965}.


\section{Previous $^{39}$Ar half-life measurements}

Until now, not many measured values of $^{39}$Ar half-life using different techniques have been published. How to assess the measurement uncertainties is a challenging problem in physics for different research teams, especially for experiments that were done a long time ago~\cite{zeldesHalflifeMassAssignment1952,stoennerHalflivesArgon37Argon391965}. On the other hand, even if  only statistical uncertainty is indicated~\cite{stoennerHalflivesArgon37Argon391965}, it could be viewed as the lower limit on the true errors~\cite{gott2001median}. 

The half-life values of the $^{39}$Ar isotope were carefully determined previously using an accelerator mass spectrometry (AMS) technique and activity measurements~\cite{zeldesHalflifeMassAssignment1952,stoennerHalflivesArgon37Argon391965}. The current accepted half-life value of the $^{39}$Ar isotope is $268 \pm 8$~years~\cite{NSR2018CH17}. It was changed from the $269 \pm 3$~years value in a previous compilation of nuclear data for atomic mass $A=39$~\cite{SINGH2006225,SINGH20111391} originally reported in Stoenner et al. (1965)~\cite{stoennerHalflivesArgon37Argon391965}, with statistical error representing the standard deviation of all 10 AMS measurements shown in the manuscript~\cite{stoennerHalflivesArgon37Argon391965}. Note that the reported result in Stoenner et al. (1965)~\cite{stoennerHalflivesArgon37Argon391965} is an updated half-life value of $^{39}$Ar isotope that was first determined as $T_{1/2}=325$~years in an earlier publication by Stoenner et al. (1960)~\cite{stoenner1960meteorites}. 

Holden (1990)~\cite{holden1990total} found that the half-life value of the $^{39}$Ar isotope was underestimated and needed to be updated, which resulted in a recommended value of  $T_{1/2}=268 \pm 8$~years; however, it was not included in the previous compilation of nuclear data for $A=39$~\cite{SINGH2006225,SINGH20111391}. It is fascinating to note that both half-life values of the $^{39}$Ar isotope $T_{1/2}=269(3)$~\cite{SINGH2006225,SINGH20111391} and $T_{1/2}=268(8)$~\cite{NSR2018CH17} are still used in recent publications (see for example Saldanha et al. (2019)~\cite{saldanha2019cosmogenic} and Zhang \& Mei (2022)~\cite{zhang2022evaluation}).

\begin{table}[t]
\centering
\caption{All  half-life measurements to date for $^{39}$Ar used to estimate weighted mean value (see Equation \ref{Eq:ar39HLwm}). The first 10 values are shown in Figure~\ref{fig:Ar39HL}. }
		\begin{tabular}{@{}cccc@{}}
		\toprule			
		Measurement &
		\begin{tabular}{c}
			Half-life, \\
			years
		\end{tabular}
		& Method & Ref. \\
		\midrule
		1 & 270   & relative to $^{37}$Ar & \cite{stoennerHalflivesArgon37Argon391965} \\
		2 & 258   & relative to $^{37}$Ar & \cite{stoennerHalflivesArgon37Argon391965} \\
		3 & 264   & relative to $^{37}$Ar & \cite{stoennerHalflivesArgon37Argon391965} \\
		4 & 288   & relative to $^{37}$Ar & \cite{stoennerHalflivesArgon37Argon391965} \\
		5 & 269   & relative to $^{37}$Ar & \cite{stoennerHalflivesArgon37Argon391965} \\
		6 & 274   & relative to $^{37}$Ar & \cite{stoennerHalflivesArgon37Argon391965} \\
		7 & 265   & relative to $^{37}$Ar & \cite{stoennerHalflivesArgon37Argon391965} \\
		8 & 272   & relative to $^{37}$Ar & \cite{stoennerHalflivesArgon37Argon391965} \\
		9 & 272   & relative to $^{37}$Ar & \cite{stoennerHalflivesArgon37Argon391965} \\
		10 & 253   & relative to $^{37}$Ar & \cite{stoennerHalflivesArgon37Argon391965} \\
		11 & 240   & AMS + activity &\cite{zeldesHalflifeMassAssignment1952} \\
		12 & 265   & AMS + activity &\cite{zeldesHalflifeMassAssignment1952} \\
		13 & 290   & AMS + activity &\cite{zeldesHalflifeMassAssignment1952} \\
		\botrule
	\end{tabular}%
\label{tab:Ar39HL}%
\end{table}%

\begin{figure*}[!htbp]
\centering
\includegraphics[width=0.9\textwidth]{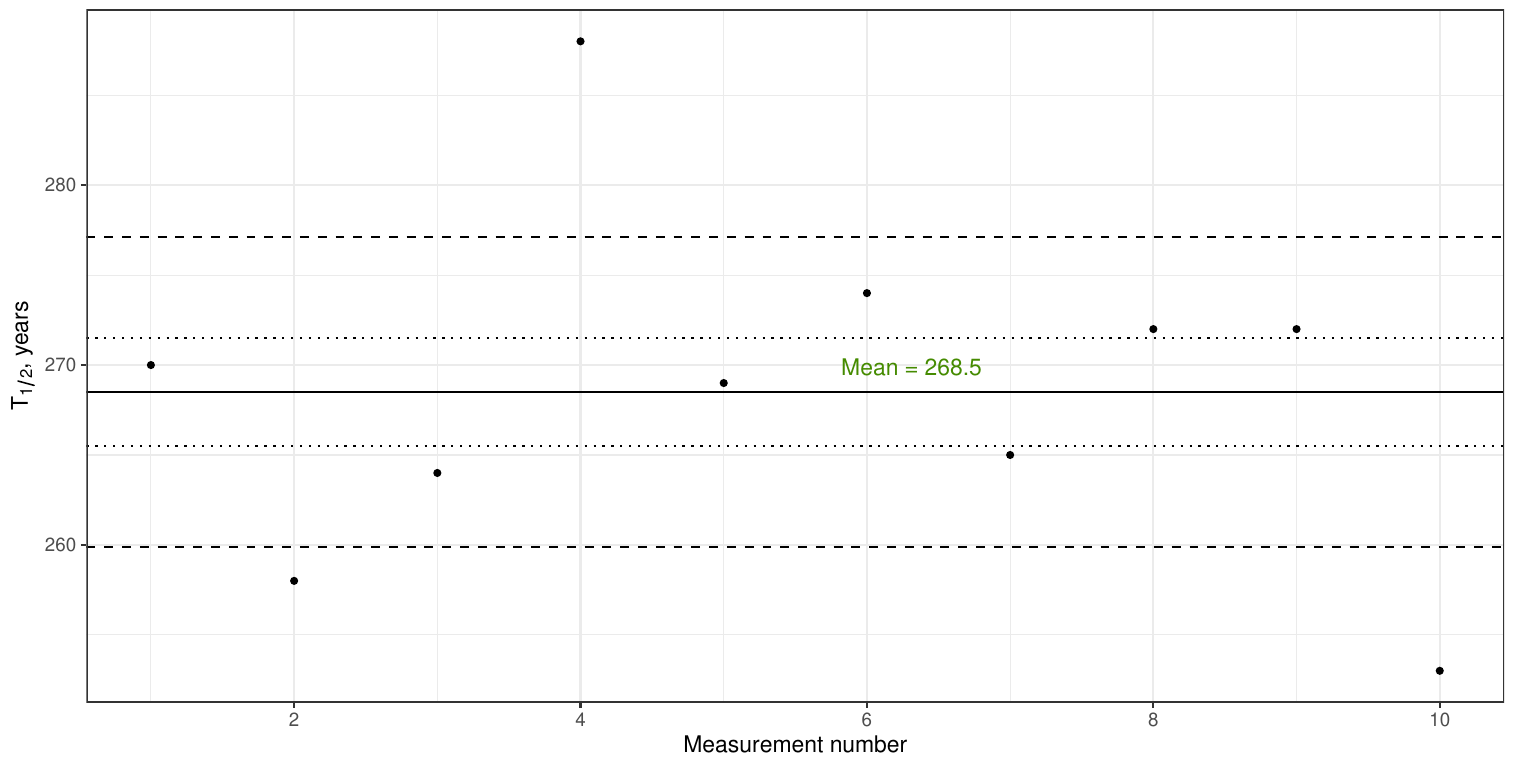}
\caption{The measured half-life of $^{39}$Ar as a function of measurement number. All 10 half-life data were taken from Table~1 in Stoenner et al. (1965)~\cite{stoennerHalflivesArgon37Argon391965}. The solid horizontal line represent the mean value for the $^{39}$Ar half-life, the dotted line represents statistical error as reported in Stoenner et al. (1965)~\cite{stoennerHalflivesArgon37Argon391965}, and the dashed line represents total error (see Equation \ref{Eq:ar39HL_corr}). }
\label{fig:Ar39HL}
\end{figure*}

For some reason the half-life error of 3~years  reported in Stoenner et al. (1965)~\cite{stoennerHalflivesArgon37Argon391965} does not include the systematic error of 3\% discussed in the original manuscript \cite{stoennerHalflivesArgon37Argon391965} and was adopted in the previous compilation in nuclear data sheets for atomic mass $ A=39 $~\cite{SINGH2006225,SINGH20111391}. In other words, based on the mean value for half-life of $^{39}$Ar of 269~years the systematic error is $\pm 8$~years. Therefore, the half-life value of $^{39}$Ar reported in Stoenner et al. (1965)~\cite{stoennerHalflivesArgon37Argon391965} should be taken as $T_{1/2} = 269 \pm 3 (\text{stat}) \pm 8 ({\text{sys}}) $~years or
\begin{equation}
	T_{1/2} = 269 \pm 9 \, \text{years}
	\label{Eq:ar39HL_corr}
\end{equation}
(i.e., the quadrature sum of the statistical and systematic uncertainties), where the total error is  dominated by the systematic error. The result in Equation~\ref{Eq:ar39HL_corr} also suggests that the error in Stoenner et al. (1965)~\cite{stoennerHalflivesArgon37Argon391965} may be  underestimated. Figure~\ref{fig:Ar39HL} shows the half-life measurements for  $^{39}$Ar isotope taken from Table~1 in Stoenner et al. (1965)~\cite{stoennerHalflivesArgon37Argon391965} (or see the first 10 values in Table~\ref{tab:Ar39HL}). 

In general, the total measurement error should represent 68\% of all half-life measurements. A ``rule of thumb'' for the confidence interval of 68\% indicates that approximately two-thirds of all $^{39}$Ar half-life measurements should be within the range indicated by dashed lines in Figure~\ref{fig:Ar39HL}. Indeed, we observe 7 out of 10 measurements in the range shown by two dashed horizontal lines for the $^{39}$Ar half-life value reported in Equation~\ref{Eq:ar39HL_corr}. Moreover, the statistical errors indicated by horizontal dotted lines contain only 2 measurements and cannot be representative of the total error for the half-life of $^{39}$Ar as was adopted in Singh \& Cameron (2006)~\cite{SINGH2006225}. From Figure~\ref{fig:Ar39HL} it is obvious that the systematic error of 3\% is adequate. 

The other half-life measurements for $^{39}$Ar isotope using mass spectrometry as we mentioned previously were reported in Zeldes et al. (1952)~\cite{zeldesHalflifeMassAssignment1952}. The presented result of $265 \pm 30$~years for $^{39}$Ar half-life measurement is completely dominated by the statistical error resulting from the averaging of three specific activity measurements reported in Table~II in Zeldes et al. (1952)~\cite{zeldesHalflifeMassAssignment1952}. Those three individual results for $^{39}$Ar half-life measurements could be deduced using Equation~\ref{Eq:SA}, the specific activity for atoms of mass 39 (or the ``disintegration per second'' value), and the total number of atoms of mass 39. All measured data were reported in the same table of the original manuscript. 

Zeldes et al. (1952)~\cite{zeldesHalflifeMassAssignment1952} did not indicate the systematic error of the half-life value, but the standard deviation of all three reported measurements was indicated to be $\pm 25$~years. Assuming that the indicated error value of $\pm30$~years represents the total error for the $^{39}$Ar isotope half-life of $T_{1/2} = 265 \pm 30$~years, one can estimate the systematic error for the half-life or  $T_{1/2} = 265 \pm 25 (\text{stat}) \pm 17 ({\text{sys}}) $~years. The ``systematic'' error of $\pm17$~years here corresponds to the 6.5\% correction discussed in the original manuscript \cite{zeldesHalflifeMassAssignment1952}.

The weighted mean ($\bar{x}_w$) is typically calculated using the following formula~\cite{Workman:2022ynf}:
\begin{equation}
\bar{x}_w = \frac{\sum_{i=1}^{n} \left( x_i \cdot \frac{1}{{\sigma_i}^2} \right)}{\sum_{i=1}^{n} \frac{1}{{\sigma_i}^2}}.
\label{eq:w_mean}
\end{equation}
In this formula:  $x_i$ stands for the data points,  $\sigma_i$  represents the associated errors, and  $n$  denotes the total number of data points. The formula for determining the error of the weighted mean  ($\sigma_{\bar{x}_w}$)  is as follows:
\begin{equation}
\sigma_{\bar{x}_w} = \frac{1}{\sqrt{\sum_{i=1}^{n} \frac{1}{{\sigma_i}^2}}}.
\label{eq:w_mean_err}
\end{equation}
Here: $\sigma_{\bar{x}_w}$  signifies the error related to the weighted mean and $\sigma_i$ still represents the individual errors. These equations describe how the weighted mean and its error are computed based on the given data and errors.

One can estimate a weighted mean (see Equation~\ref{eq:w_mean}) half-life and an error of the weighted mean (see Equation~\ref{eq:w_mean_err}) for the $^{39}$Ar isotope by using the previous discussion and by averaging the $^{39}$Ar half-life values cited in the literature~\cite{zeldesHalflifeMassAssignment1952} and re-calculated $^{39}$Ar half-life~\cite{stoennerHalflivesArgon37Argon391965} (in other words the value from Equation~\ref{Eq:ar39HL_corr}), which results in  
\begin{equation}
	T_{1/2} = 268.7 \pm 8.6 \, \text{years.}
	\label{Eq:ar39HLwm}
\end{equation}
Quoted uncertainties are given at the 68\% confidence level, which corresponds to the [260.1, 277.3] confidence interval, whereas the 95\% confidence interval for all data is [251.5, 285.9]. 

This value is close to the $^{39}$Ar half-life of  $T_{1/2} = 268 \pm 8  $~years that was reported by Holden (1990)~\cite{holden1990total} after re-analysis of the original data from Stoenner et al. (1965)~\cite{stoennerHalflivesArgon37Argon391965}. Holden (1990)~\cite{holden1990total} recommends increasing the total error in the original half-life data as it is underestimated by a factor of 3, and also recommends using a half-life value of $^{37}$Ar ($T_{1/2} = 35.02 \pm 0.05 $~day~\cite{kishore1975cl}) that is more precise than the value of $T_{1/2} = 35.1 \pm 0.1 $~day used in original paper~\cite{stoennerHalflivesArgon37Argon391965}) to deduce $^{39}$Ar half-life. However, the recommendation was not taken into account in the previously adopted half-life of $^{39}$Ar in nuclear data sheets~\cite{SINGH2006225}. Table~\ref{tab:Ar39HL} summarize all known half-life measurements to date for the $^{39}$Ar isotope to the best of the author's knowledge.

\section{MFV approach for robust estimate of $^{39}$Ar half-life}

Hashimoto et al. (2001) developed a novel implantation method~\cite{hashimoto2001half} by which the isotope in question is produced as a radioactive beam and implanted into a stopper, and its specific activity is measured after the irradiation. Combined with the AMS technique, this raised interest in the preparation of $^{39}$Ar samples to investigate its half-life~\cite{fulopa2005production}. However, this required a new measurement that is quite expensive and time-consuming. 

The alternative could be the application of the MFV approach that was recently used for a robust estimate of neutron lifetime (Zhang et al., 2022)~\cite{zhang2022mfv}. An improved, iteratively re‐weighted factor analysis procedure based on the MFV approach was used to interpret engineering geophysical sounding logs in shallow unsaturated sediments \cite{szaboMostFrequentValuebased2018}. Moreover, the MFV approach was applied to groundwater modeling as a robust and effective geostatistical method \cite{szucsApplicabilityMostFrequent2006}. The MFV algorithm is  based on the principle of minimization of the information lost, and was applied to determine the Hubble constant, regardless of  the Gaussian or non-Gaussian distributions  (Zhang, 2018)~\cite{zhang2018most}. In addition, MFV was applied to another nuclear astrophysical challenge recently, namely the lithium abundance problem (Zhang, 2017)~\cite{zhang2017most}.

The advantage of the MFV statistical technique~\cite{steinerMostFrequentValue1988,steinerMostFrequentValue1991,steinerOptimumMethodsStatistics1997,szucsApplicabilityMostFrequent2006} is that MFV is found to be independent of the statistical distribution of input data (for example, for non-normally distributed data or in the presence of outliers). In addition, the MFV is recommended to give a more robust estimate, which calculates the weighted average of the evaluation points in a statistically highly efficient algorithm. The MFV technique provides optimal weight coefficients for any given dataset due to its  capability to automatically calculate the scale parameter during the iterative process. 

The explicit equation of the iterations for the most frequent value is given as
\begin{equation}
	MFV_{j+1} = \frac{ \sum_{i=1}^{N} x_i \cdot \frac{\varepsilon^2_j}{\varepsilon^2_j + \left( x_i - MFV_j \right)^2 } }{ \sum_{i=1}^{N} \frac{\varepsilon^2_j}{\varepsilon^2_j + \left( x_i - MFV_j \right)^2} } \, ,
	\label{Eq:MFV}
\end{equation}
where $ MFV_{j+1} $ is the MFV iteration process to obtain the $ (j+1) $-th step, $ x_i $ is the $ i $-th element in a dataset, $ N $ is the total number of elements in dataset, and $\varepsilon_j$ is the dihesion (also known as Steiner's scale factor~\cite{szegedi2014use}). The initial value for the MVF iteration could be taken as a mean value for all data in a sample (or $ MFV_{(0)} = \frac{1}{N} \sum_{i=1}^{N} x_i$).  

The equation of the iterations for the dihesion $ \varepsilon_{j+1} $ is given as
\begin{equation}
	\varepsilon^2_{j+1} = 3 \cdot \frac{ \sum_{i=1}^{N}  \frac{(  x_i - MFV_j)^2 }{ \left( \varepsilon^2_j + \left( x_i - MFV_j \right)^2 \right)^2 } }{ \sum_{i=1}^{N}  \frac{ 1 }{ \left( \varepsilon^2_j + \left( x_i - MFV_j \right)^2 \right)^2 }  } \, ,
	\label{Eq:dihesion}
\end{equation}
where the value of $\varepsilon_j$ is estimated in the previous iteration step. The initial value of $\varepsilon_{(0)}$ can be chosen as $ \varepsilon_{(0)} = \frac{\sqrt{3}}{2} \cdot (x_{max} - x_{min}) $, where $ x_{max} $ and $ x_{min} $ are the maximum and minimum values in the entire dataset used for MFV estimation. The iteration threshold value in Equations~\ref{Eq:MFV} and \ref{Eq:dihesion} could be chosen to be an arbitrary small value (for example, $ 10^{-5} $). In other words, the iteration threshold value is the limit that needs to be reached through iterations on the next difference $ \left| MFV_{j+1} - MFV_{j} \right| $ in Equation~\ref{Eq:MFV}. 
Steiner (1988)~\cite{steinerMostFrequentValue1988} showed that the iteration given by Equations~\ref{Eq:MFV} and \ref{Eq:dihesion} approximates a value characteristic of the concentration of data, and is therefore called the most frequent value \cite{steinerOptimumMethodsStatistics1997}.

\begin{figure*}[!htbp]
	\centering
	\includegraphics[width=0.9\textwidth]{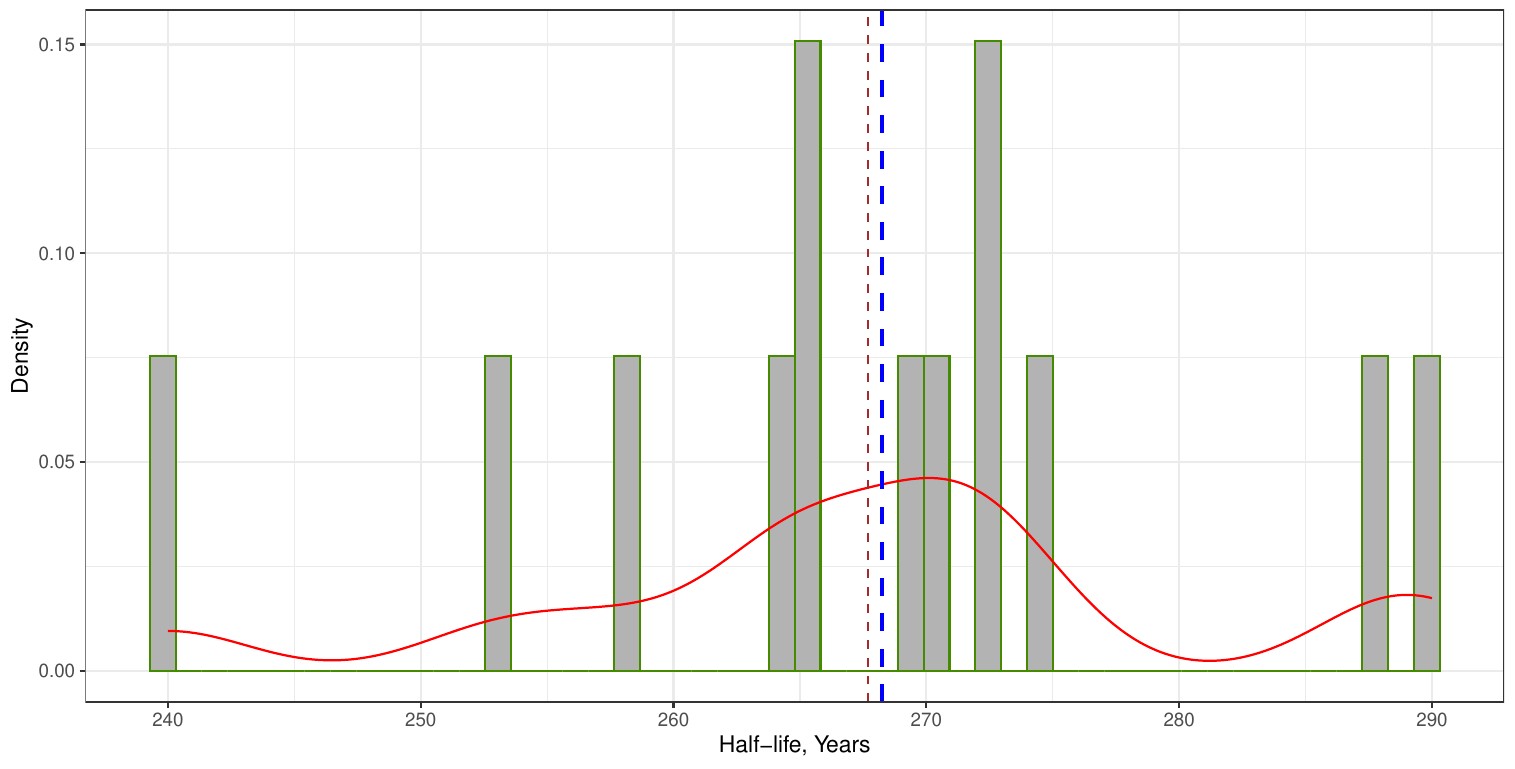}
	\caption{Histogram and probability density (solid line) of $^{39}$Ar half-life values (see Table~\ref{tab:Ar39HL}). The vertical thin dashed line is the mean value and the vertical thick dashed line is the MFV.
	}
	\label{fig:Ar39his}
\end{figure*}

For a symmetrical distribution (for example on normally distributed data) a simple formula was provided by Cserny\a'ak and Steiner \cite{steinerOptimumMethodsStatistics1997} for variance $ \sigma_{\text{MFV}}$  (i.e., if $ MFV $ is calculated as the most frequent value following Equation~\ref{Eq:MFV}) as
\begin{equation}
	\sigma_{MFV} = \frac{\varepsilon}{\sqrt{n_{\text{eff}}}} \, ,
	\label{Eq:sigma}
\end{equation}
where $ \varepsilon $ is the dihesion (see Equation~\ref{Eq:dihesion}) and $ n_{\text{eff}} $ is the effective number of data. The effective number is defined as
\begin{equation}
	n_{\text{eff}} = \sum_{i=1}^{N} \frac{\varepsilon^2}{\varepsilon^2 + \left( x_i - MFV \right)^2} \, .
	\label{Eq:neff}
\end{equation}

The mean of all half-life measurements of $^{39}$Ar isotope taken from Table~\ref{tab:Ar39HL} was found to be $267.7$~years, whereas the MFV of all half-life measurements of $^{39}$Ar isotope was found to be $268.2$~years. Both values are shown as vertical lines in Figure~\ref{fig:Ar39his}. To summarize, historically there were only two half-life measurements for the $^{39}$Ar isotope: one with a narrow confidence interval~\cite{stoennerHalflivesArgon37Argon391965}, and another with quite a large confidence interval~\cite{zeldesHalflifeMassAssignment1952}. All historically reported  confidence intervals overlap with each other; however, the narrow one is too optimistic, as it does not include systematic errors (see Figure~\ref{fig:Ar39HL} and previous discussion in the text). As  mentioned before, the systematic errors are discussed in the paper, but only statistical errors were used to define confidence interval for  half-life measurements of the $^{39}$Ar isotope. 

In principle, one expensive and time-consuming option to obtain a confidence interval would be to replicate the half-life measurement experiment several times. If one repeats the experiment several times, then one can keep track of each half-life value and end up with a larger set of half-life data that could be used to estimate the $^{39}$Ar half-life value and its confidence interval. However, as mentioned earlier, repeating the half-life experiment several times is both expensive and time-consuming. 

Suppose one has a set of data that represent true $^{39}$Ar half-life  measurements (see Table~\ref{tab:Ar39HL}), and one needs to find a confidence interval based on statistical methods (for example MFV). Instead of replicating the half-life experiment several times one can use a bootstrap approach~\cite{efron1994introduction,davisonBootstrapMethodsTheir1997}. An additional advantage of bootstrapping is that bootstrap methods could be applied both when there is a well-defined  probability model for data and when there is not~\cite{davisonBootstrapMethodsTheir1997}. So let's use the bootstrap technique to get a better sense of which confidence interval more likely represents $^{39}$Ar half-life  measurements. Moreover, one can use a robust estimate based on MFV statistics that could be applied to non-normally distributed data. 

First, one can create a new data sample from the 13 historical $^{39}$Ar half-life  measurements reported in Table~\ref{tab:Ar39HL} by choosing randomly from the distribution of half-life values with replacement. One can randomly select the same half-life value more than once; this process of randomly selecting data and allowing for duplicates is called sampling with replacement. As the original dataset contains only 13 half-life values the new random sample that is created with replacement also contains 13 elements and is called a bootstrapped dataset. In addition, the published $^{39}$Ar half-life values do not contain observational errors of every measurement shown in Table~\ref{tab:Ar39HL}, and therefore the bootstrap method to evaluate confidence interval is appropriate~\cite{efron1994introduction,davisonBootstrapMethodsTheir1997}. 

The new dataset is different from the true half-life value dataset (see Table~\ref{tab:Ar39HL}), and therefore one can apply MFV statistics (see Equations~\ref{Eq:MFV} and \ref{Eq:dihesion}) to it and obtain a new MFV value. Finally, repeating this process $ B $ times (usually 1,000--3,000 times) generates the distribution of the MFV statistic, and the process is called bootstrapping. From the generated MFV distribution one can calculate  confidence intervals at 68\% and 95\% confidence levels. Additional information on bootstrap and bootstrapping techniques is provided in detail elsewhere~\cite{efron1994introduction,davisonBootstrapMethodsTheir1997}. 

As mentioned previously, the MFV technique and confidence interval bootstrapping had been used for a robust analysis of neutron lifetime measurements (Zhang et al., 2022~\cite{zhang2022mfv}). We used MFV neutron lifetime data from the original manuscript (Zhang et al., 2022~\cite{zhang2022mfv}) as a test ground for the MFV algorithm and obtained the same result.

Use of the bootstrapping technique for $^{39}$Ar isotope half-life values presented in Table~\ref{tab:Ar39HL} and MFV statistics at the 68\% confidence level resulted in a confidence interval for all measurements of [265.3, 271.3], whereas the 95\% confidence interval for all data is [262.3, 274.3]. In other words, an estimation of the $^{39}$Ar isotope half-life based on MFV statistics and a bootstrapping technique results in 
\begin{equation}
	T_{1/2}({\text MFV}) = 268.2^{+3.1}_{-2.9} \, \, \text{years,}
	\label{Eq:ar39HLmfv}
\end{equation}
with a $1 \sigma$ range of [265.3, 271.3] with uncertainty corresponding to the 68\% confidence level. The MFV $^{39}$Ar isotope half-life result is closed to the value reported in Holden (1990)~\cite{holden1990total}; however, here we have not used a more precise half-life value of $^{37}$Ar. One can estimate the MFV variance for the $^{39}$Ar isotope half-life using Equation~\ref{Eq:sigma}, which results in $ \sigma_{MFV} = 2.5 $~years. The variance is close to the bootstrap errors shown in Equation~\ref{Eq:ar39HLmfv}. The advantage of  bootstrapping versus MFV variance is that no assumption is made for a distribution of the data.

\begin{figure*}[!htbp]
\centering
\includegraphics[width=0.9\textwidth]{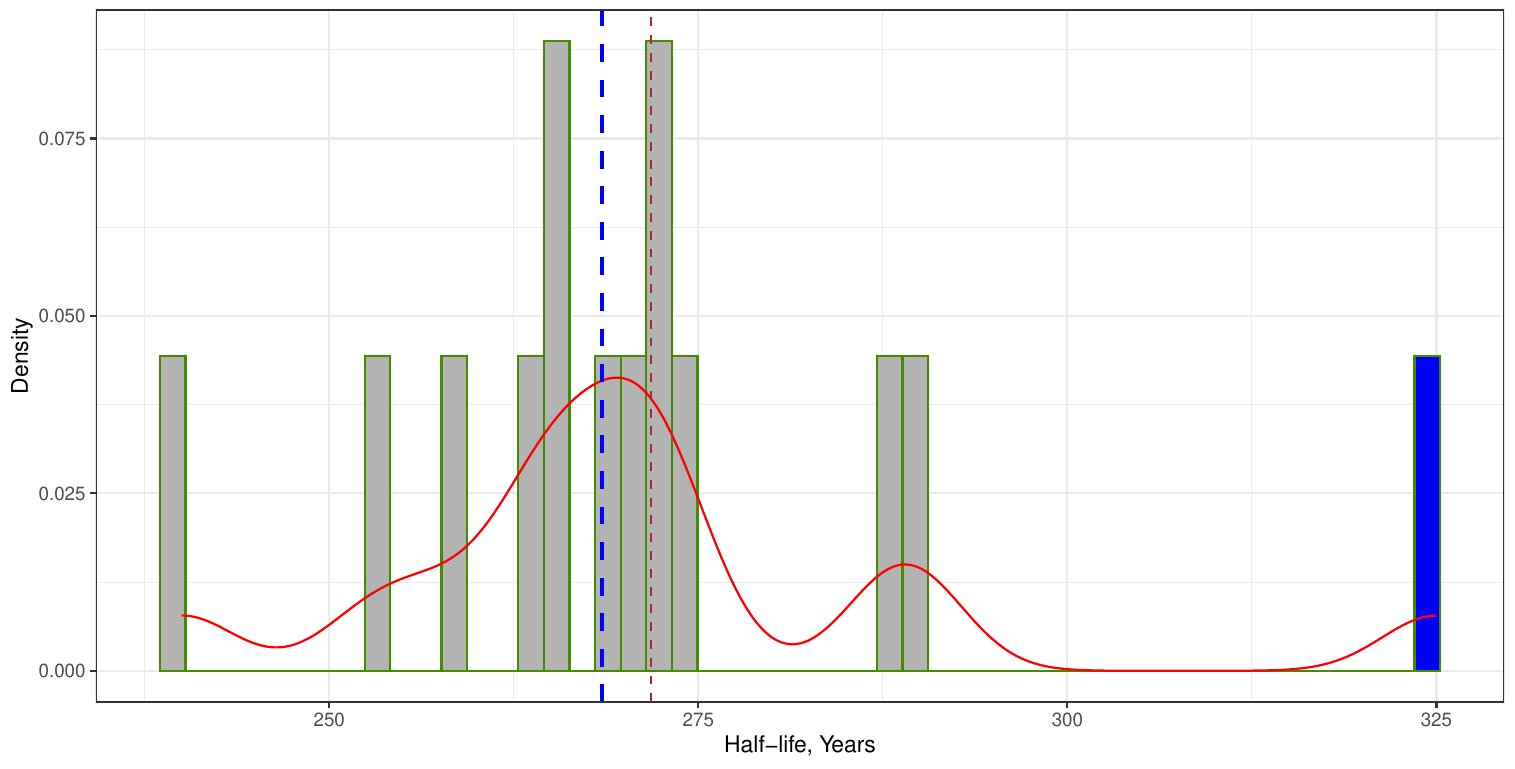}
\caption{Histogram and probability density (solid line) of $^{39}$Ar half-life values (see  Table~\ref{tab:Ar39HL})   with addition of 325~years (presence of outlier). The vertical thin dashed line is the mean value and vertical thick dashed line is the MFV.  We've chosen to highlight the particular value of 325 years by using a darker color. This helps set it apart from the measurements of $^{39}$Ar half-life listed in Table~\ref{tab:Ar39HL}. 
}
\label{fig:Ar39_w325}
\end{figure*}

It has been shown that the MFV approach is a robust estimate and  tolerates the presence of outliers in the dataset~\cite{steinerMostFrequentValue1988,steinerMostFrequentValue1991,steinerOptimumMethodsStatistics1997}. As an exercise one can check this by adding a data point to the  $^{39}$Ar half-life measurements collected in Table~\ref{tab:Ar39HL}. The additional value was taken from the preliminary result for $^{39}$Ar half-life of $T_{1/2}=325$~years in Stoenner et al. (1960)~\cite{stoenner1960meteorites}. The newly added data point in the dataset changed the estimation of $^{39}$Ar isotope half-life based on MFV statistics and bootstrapping technique to  $T_{1/2}({\text MFV}^*) = 268.5^{+3.2}_{-3.4} \, \text{years,}$ with a $1 \sigma$ range of [265.1, 271.7] with uncertainty corresponding to the 68\% confidence level (a 2$\sigma$ range is [261.7, 275.0] corresponding to the 95\% confidence level). The mean value for the new dataset is 271.8~years (higher than the mean value of 267.7~years for data presented in Table~\ref{tab:Ar39HL}). It is clear that the value of the  mean is strongly influenced by the half-life value of 325~years (see Figure~\ref{fig:Ar39_w325}).

In summary, the problem of  accuracy in the half-life of the $^{39}$Ar isotope is one of the important challenges of particle physics. The accuracy is especially crucial in practical applications for $ ^{40} $Ar/$ ^{39} $Ar dating methods  in the field of isotope geochronology \cite{renne1998intercalibration}. 
Modern measurements of $ ^{40} $Ar:$ ^{38} $Ar:$ ^{36} $Ar by Lee et al. (2006)~\cite{lee2006redetermination} provide  more accurate reference corrections for $ ^{40} $Ar/$ ^{39} $Ar geochronology~\cite{renneIsotopicCompositionAtmospheric2009}. Moreover, the accurate $^{39}$Ar half-life is used to constrain the age distribution of groundwater~\cite{corchoalvaradoConstrainingAgeDistribution2007}, for complementary hydrographic and nutrient data studies~\cite{schlitzerMeridional14C39Ar1985} and, as was mentioned previously, as a probe for groundwater dating~\cite{loosliUse39Ar14C1980,LEHMANN1997727}. Using the MFV approach in combination with bootstrapping allows a robust estimate of the $^{39}$Ar half-life and its accuracy (see Equation~\ref{Eq:ar39HLmfv}) without the expensive and time-consuming option of replicating half-life measurements using the AMS technique as done previously \cite{zeldesHalflifeMassAssignment1952,stoennerHalflivesArgon37Argon391965}.

\section{Specific Atmospheric $^{39}$Ar Activity}

All radionuclides are uniquely identified by three characteristics: the type of radiation, the energy of the radiation, and the rate at which the radioactive decay occurs. The concentration of radioactivity, or the relationship between the mass of radioactive material and the activity, is called the specific activity. If the radioactive material consists of a single radioisotope, the mass of radioactive material represents the mass of that particular isotope. The specific activity is the number of becquerels (or curies) per unit mass or volume~\cite{cemberIntroductionHealthPhysics2009}. If $\lambda$ is the decay constant for the radionuclide under consideration in units of reciprocal seconds, then the number of decays for the radioactive isotope per second and, hence, the number of becquerels in $ N $ atoms, is simply given by $ \lambda \cdot N $. The activity per unit weight or the specific activity $ SA $, therefore, is
\begin{equation}
	SA = \lambda \cdot N = \frac{ \lambda \cdot N_A } { A },
	\label{Eq:SA}
\end{equation}
where $ A $ is the atomic weight of the nuclide, and $ N_A $ is the Avogadro’s number of atoms for the radionuclide under consideration. The decay constant $ \lambda $ of an unstable radioactive nucleus is obtained as $ \lambda = \frac{\text{ln}(2)}{ T_{1/2}} $~\cite{grupenIntroductionRadiationProtection2010}. Therefore, the specific activity for radioactive isotope relates to the half-life of the same isotope as 
\begin{equation}
	SA = \frac{ \text{ln}(2) \cdot N_A } { T_{1/2} \cdot A }.
	\label{Eq:SA_T}
\end{equation}
An accurate half-life determination that uses data from specific activity measurements is often  used for long-life radionuclides (see for example Brown et al., (1981) and Kossert \& G{\"u}nther (2004)~\cite{brownRedeterminationHalflife239Pu1981,kossertLSCMeasurementsHalflife2004}). The specific activity of a particular isotope is very useful when dealing with the classification of nuclear waste.

As mentioned in the introduction radioargon is produced in the atmosphere. The half-life of $^{39}$Ar is long  compared with  mixing time in the atmosphere. The $^{39}$Ar isotope in the atmosphere should have a homogeneous distribution and the decay should be in equilibrium with its production rate.
Recent work on reconstruction of the atmospheric $^{39}$Ar/Ar history shows that the temporal variation of the atmospheric $^{39}$Ar in the past 2,500 years has changed as much as 17\% in that period~\cite{guReconstructionAtmospheric39Ar2021}. Moreover, it has been shown \cite{guReconstructionAtmospheric39Ar2021} that the anthropogenic contribution to the atmospheric $^{39}$Ar in the past 60~years is less than 15\%. Taking this into account, one should conclude that the radioactive isotope $^{39}$Ar is constant over a short period (relative to 2,500~years) and that it is an ideal tracer for dating applications (Loosli, 1983~\cite{loosliDatingMethod39Ar1983}).  

The other cosmic-ray-produced radionuclides, such as the $ ^{10} $Be and $ ^{14} $C that are stored in polar ice cores and tree rings, offer the unique opportunity to reconstruct the history of cosmic radiation and solar activity over 9,400 years~\cite{steinhilber400YearsCosmic2012}. To compare the 9,400-year  record of cosmic ray intensity with the most recent years, neutron monitoring and ionization chamber data were used to calculate cosmic ray intensity. The modern 22-year average of cosmic-ray intensity based on $ ^{10} $Be data shows no variation \cite{steinhilber400YearsCosmic2012}.

Recent measurements of specific atmospheric $^{39}$Ar activity made in various underground laboratories~\cite{adhikariPrecisionMeasurementSpecific2023,adhikariPrecisionMeasurementSpecific2023b,benetti2007measurement,calvoBackgroundsPulseShape2018} show a value that is very close to the old specific atmospheric $^{39}$Ar activity reported in Loosli et al. (1970)~\cite{loosliArgon37Argon1970} and to the activity of argon samples extracted in 1940 and between 1959 and 1960. $^{39}$Ar half-life uncertainties in specific atmospheric $^{39}$Ar activity have magnified effects on specific activities reported previously because of their  appearance in  Equation~\ref{Eq:SA_T}.

\begin{figure*}[!htbp]
	\centering
	\includegraphics[width=0.9\textwidth]{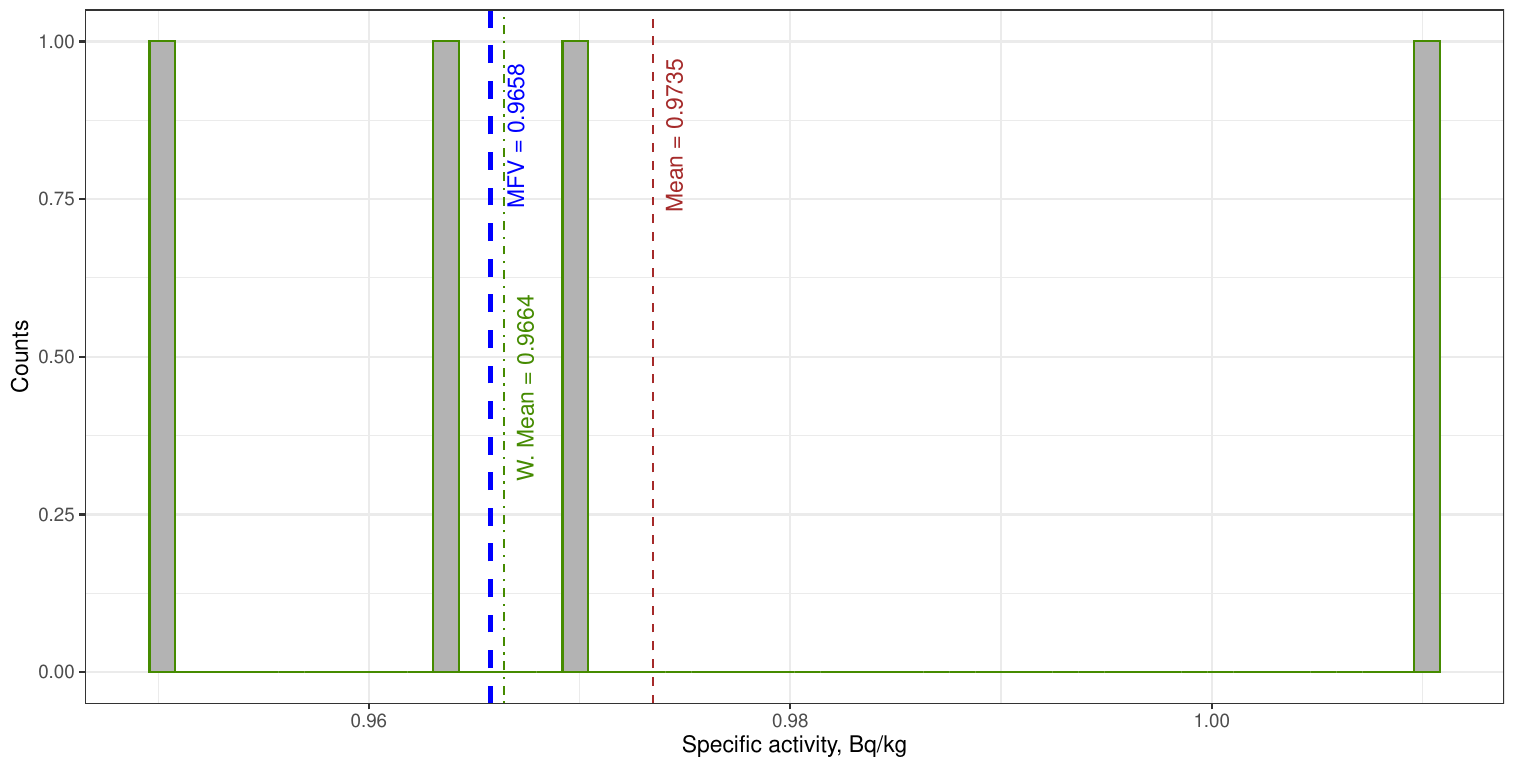}
	\caption{Histogram of $^{39}$Ar specific activity values (see Table~\ref{tab:SA_Ar39}). The vertical dotted line is the mean value and the vertical dashed line is the MFV. The vertical dot-dashed line represents the weighted mean value (as defined in Equation~\ref{eq:w_mean}). 
	}
	\label{fig:SA_Ar39his}
\end{figure*}

The absolute specific atmospheric $^{39}$Ar activity is not needed for dating, but it was nevertheless determined in tropospheric argon samples using several proportional counters and several calibration sources. An average value of $ 0.107 \pm 0.004 $ dpm/l was obtained (Loosli, 1983~\cite{loosliDatingMethod39Ar1983}). The techniques for measuring the activity of $^{39}$Ar in the environment used by Loosli is provided in Forster et al., (1992)~\cite{forsterCurrentTechniquesMeasuring1992}. The cosmogenic $^{39}$Ar concentration in the atmosphere defines the atmospheric reference and is also known as the 100 percent modern argon (pmAr) concentration (see for example Mace et al., (2017) and Ritterbusch et al., (2014)~\cite{mace2017methods,ritterbuschGroundwaterDatingAtom2014}).

Based on the absolute specific atmospheric $^{39}$Ar activity provided by Loosli (1983)~\cite{loosliDatingMethod39Ar1983} one can deduce the fraction of $^{39}$Ar in natural argon gas. For that one can find the total number of  $^{39}$Ar atoms using Equation~\ref{Eq:SA} and the $^{39}$Ar half-life from Equation~\ref{Eq:ar39HLmfv}. Therefore, the number of $^{39}$Ar atoms would be $ N_{^{39}\text{Ar}}= 2.184(85)\cdot 10^{7}$.

As the abundance of the $^{40}$Ar isotope is 99.6\%~\cite{lee2006redetermination} one can assume that the naturally occurring argon gas consists of only  stable $^{40}$Ar isotope. For a liter of ideal gas at standard temperature and pressure conditions (STP) one can estimate the number of $^{40}$Ar atoms in it. The STP conditions are: temperature  $T = 273.15 $~K,  pressure  $p = 101.325\cdot 10^3$~Pa, and volume  $ V = 0.001 $~m$^3$. The ideal gas law states that $p \cdot V = k \cdot N \cdot T$, where $p$ is the absolute pressure of a gas, $V$ is the volume it occupies, $ N $ is the number of atoms and molecules in the gas, and $ T $ is its absolute temperature. The Boltzmann constant $ k $ is equal to $ 1.38\cdot10^{-23} $~J/K. Therefore, the number of $^{40}$Ar atoms in one liter of gas at STP would be  $N_{^{40}\text{Ar}}= 2.688 \cdot 10^{22}$ atoms.

The ratio of $N_{^{39}\text{Ar}}$ to $N_{^{40}\text{Ar}}$ provides a fraction (or abundance) of $N_{^{39}\text{Ar}}$ in natural argon. Thus, the fraction (also known as the amount-of-substance fraction~\cite{kossertLSCMeasurementsHalflife2004}) of $N_{^{39}\text{Ar}}$ in natural argon is
\begin{equation}
	f_{{^{39}\text{Ar}}/\text{Ar}} = (8.12 \pm 0.30) \cdot 10^{-16}.
	\label{Eq:ar39fr}
\end{equation}
The uncertainty in Equation~\ref{Eq:ar39fr} comes solely from the $^{39}$Ar half-life and the absolute specific atmospheric $^{39}$Ar activity reported in Loosli (1983) and Forster et al. (1992)~\cite{loosliDatingMethod39Ar1983,forsterCurrentTechniquesMeasuring1992} and corresponds to the 68\% confidence level. As one can see, in order to derive the fraction of $N_{^{39}\text{Ar}}$ in natural argon, the $^{39}$Ar half-life obtained with the MFV procedure and absolute specific atmospheric activity was used (see Equation~\ref{Eq:ar39HLmfv}). 

With an independent way to measure the fraction of $N_{^{39}\text{Ar}}$ in natural argon with a reasonable accuracy   it would be possible to derive the $^{39}$Ar half-life directly from the already measured specific atmospheric $^{39}$Ar activity \cite{loosliDatingMethod39Ar1983,benetti2007measurement,calvoBackgroundsPulseShape2018,adhikariPrecisionMeasurementSpecific2023}. See for example how the amount-of-substance fraction for $ ^{40} $K isotope was measured by Garner et al. (1975)~\cite{garner1975absolute}, and later was used for the $ ^{40} $K half-life determination~\cite{kossertLSCMeasurementsHalflife2004} in a way similar to that suggested in this work. 

The recent development of atom trap trace analysis (ATTA) based on a laser atom counting method has been applied to analyze atmospheric $^{39}$Ar~\cite{jiangAr39Detection102011}. With ATTA, 1,162 $^{39}$Ar atoms were counted in 325~hours distributed over 13 weeks~(see Table~2 in Ritterbusch et al. (2014)~\cite{ritterbuschGroundwaterDatingAtom2014}). If one  counted also $^{40}$Ar atoms using the same ATTA method, the fraction of $N_{^{39}\text{Ar}}$ in natural argon that could be derived would not rely on $^{39}$Ar half-life. Another location where the fraction of $N_{^{39}\text{Ar}}$ in natural argon could be potentially measured is the WITCH setup \cite{kozlov2008witch}  at the ISOLDE facility \cite{kugler2000isolde}.  

To illustrate this approach let us assume that one has a measured value for the fraction of $N_{^{39}\text{Ar}}$ in natural argon (let us use the value from Equation~\ref{Eq:ar39fr}) that was obtained independent of $^{39}$Ar half-life. In addition, one can use a specific atmospheric $^{39}$Ar activity obtained from underground measurements, where it was calculated by estimating the total number of $^{39}$Ar decays within a certain time~\cite{adhikariPrecisionMeasurementSpecific2023}. 

As another exercise, one can use the MFV method to determine the specific activity of atmospheric $^{39}$Ar obtained from all underground measurements summarized in Table~\ref{tab:SA_Ar39} (taken from Adhikari et al., 2023~\cite{adhikariPrecisionMeasurementSpecific2023, adhikariPrecisionMeasurementSpecific2023b}). The mean of all underground measurements of the specific activity of $^{39}$Ar isotope, as shown in Table~\ref{tab:SA_Ar39}, was found to be $0.9735$ Bq/kg$_{\text{atmAr}}$, while the MFV of all specific activity measurements of the $^{39}$Ar isotope was found to be $0.9658$ Bq/kg$_{\text{atmAr}}$. The weighted mean of all underground measurements of the specific activity of the $^{39}$Ar isotope, as shown in Table~\ref{tab:SA_Ar39}, was found to be $0.9664$ Bq/kg$_{\text{atmAr}}$, which aligns with the MFV. All these values are represented as vertical lines in Figure~\ref{fig:SA_Ar39his}. By using the MFV approach, one can obtain an estimate that aligns more closely with the expected value. Following the steps outlined in the previous section of this work the MFV for specific activity of  $^{39}$Ar from underground measurements (see Table~\ref{tab:SA_Ar39}) is 
\begin{equation}
	SA_{{^{39}\text{Ar}}/\text{Ar}}(\text{MFV}) = 0.966^{+0.010}_{-0.018} \, \, \text{Bq/kg$_{\text{atmAr} } $},
	\label{Eq:SA_Ar39mfv}
\end{equation}
with confidence interval [0.948, 0.976] corresponding to the 68\%  confidence level. The confidence interval corresponding to the 95\%  confidence level is [0.934, 0.989].

\begin{table}[t]
	\centering
	\caption{Summary of underground specific activity measurements of $^{39}$Ar by different collaborations.}
		\begin{tabular}{@{}lccc@{}}
			
			\toprule
			
			Experiment & \begin{tabular}{c}
				$SA$, \\
				Bq/kg$_{\text{atmAr} } $ \\
			\end{tabular} & \begin{tabular}{c}
				Error, \\
				Bq/kg$_{\text{atmAr} } $ \\
			\end{tabular} & Ref.\\
			\midrule
			
			DEAP-3600 & 0.964 & $ \pm $0.024 & \cite{adhikariPrecisionMeasurementSpecific2023, adhikariPrecisionMeasurementSpecific2023b} \\
			DEAP-3600 & 0.97 & $ \pm $0.03 & \cite{adhikariPrecisionMeasurementSpecific2023,adhikariPrecisionMeasurementSpecific2023b} \\
			WARP & 1.01 & $ \pm $0.08 & \cite{benetti2007measurement} \\
			ArDM  & 0.95  & $ \pm $0.05 & \cite{calvoBackgroundsPulseShape2018} \\
			\botrule
			
		\end{tabular}%
	\label{tab:SA_Ar39}%
\end{table}%

The specific activity in Equation~\ref{Eq:SA_T} has been defined for an isotope, and therefore to derive the $^{39}$Ar half-life value from it one should take into account the fraction of $N_{^{39}\text{Ar}}$ in natural argon (see Equation~\ref{Eq:ar39fr}) for the a specific atmospheric $^{39}$Ar activity obtained from underground measurements. Thus, the half-life of $^{39}$Ar is
\begin{equation}
	T_{1/2} = \frac{ \text{ln}(2) \cdot N_A \cdot f_{{^{39}\text{Ar}}/\text{Ar} } } { SA_{{^{39}\text{Ar}}/\text{Ar}}  \cdot A },
	\label{Eq:HL_SA}
\end{equation}
resulting in $T_{1/2} = 285 \pm 4$~years with the uncertainty corresponding to the 68\% confidence level. 
To determine the uncertainty in the half-life of $^{39}$Ar using Equation~\ref{Eq:HL_SA}, we applied standard error analysis methods. In simpler terms, we calculated $\Delta T_{1/2}$, which represents the uncertainty in the half-life, using the formula:
\begin{equation}
\Delta T_{1/2} = T_{1/2} \cdot \sqrt{ \left( \frac{\Delta f_{{^{39}\text{Ar}}/\text{Ar} }}{f_{{^{39}\text{Ar}}/\text{Ar} }} \right)^2 + \left( \frac{\Delta SA_{{^{39}\text{Ar}}/\text{Ar}}}{SA_{{^{39}\text{Ar}}/\text{Ar}}}   \right)^2 }. 
\end{equation}
Here, $\Delta$ denotes the respective uncertainties. This process involved taking into account the uncertainties associated with the variables used in the equation. Specifically, we treated the asymmetric uncertainties mentioned in Equation~\ref{Eq:SA_Ar39mfv} as if they were symmetric. For instance, $SA_{{^{39}\text{Ar}}/\text{Ar}}(\text{MFV}) = 0.966 \pm 0.014$ Bq/kg$_{\text{atmAr} } $.
This choice was influenced by previous studies by Wang et al. (2012) \cite{wang2012ame2012} and Barlow (2021)~\cite{barlow2021practical}. The process of arriving at this uncertainty value followed established principles of error analysis propagation, which are well-documented in references~\cite{gilmorePracticalGammaraySpectroscopy2008,hughesMeasurementsTheirUncertainties2010,knollRadiationDetectionMeasurement2010}. These references provide guidelines for a rigorous and systematic assessment of uncertainties in the final result, ensuring its reliability. 
Note that the fraction on the right-hand side of  Equation~\ref{Eq:SA_T} depends on $^{39}$Ar half-life, and therefore it will be correlated with the half-life value deduced on the left-hand side of Equation~\ref{Eq:SA_T}.

\begin{figure*}[!htbp]
\centering
\includegraphics[width=0.9\textwidth]{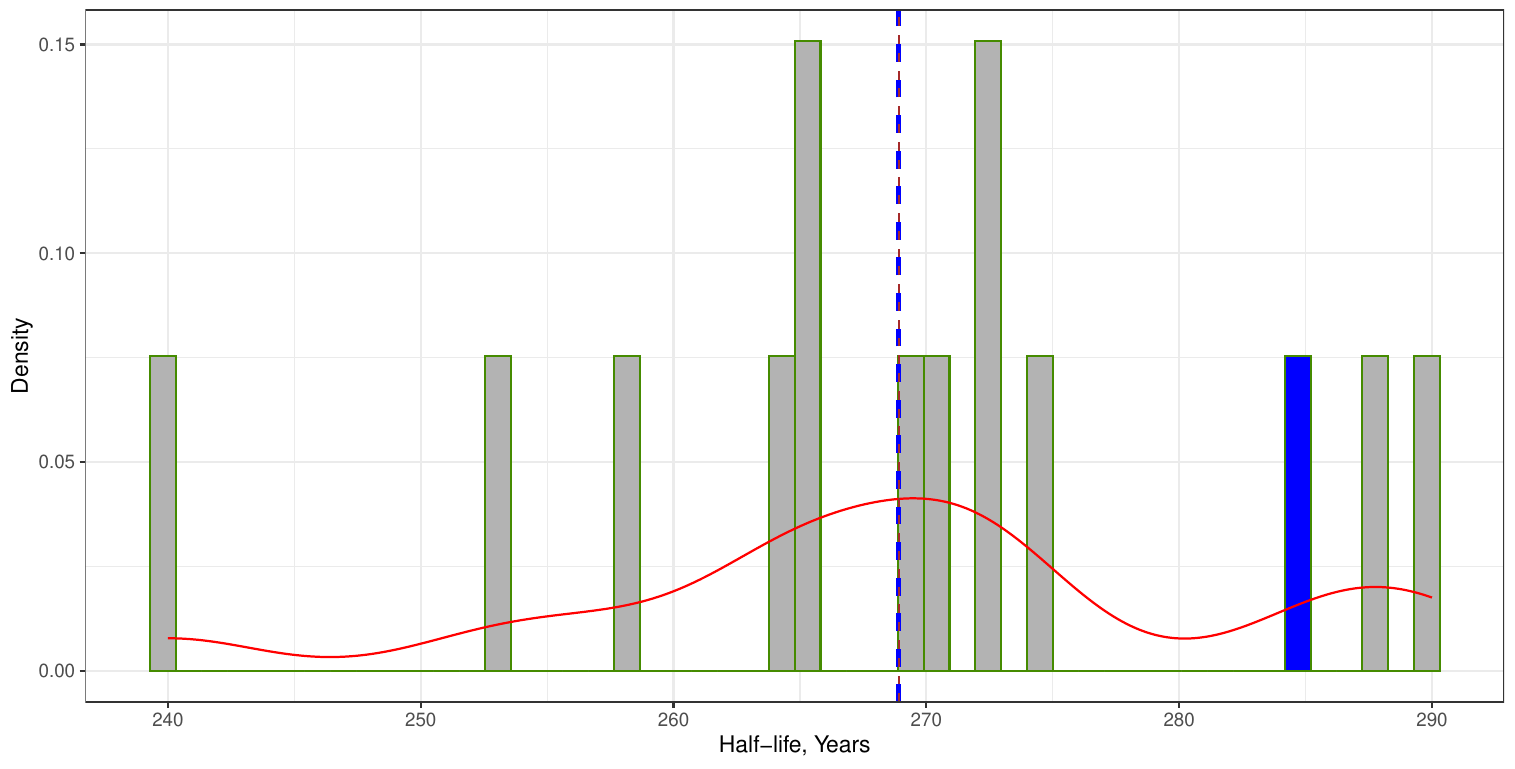}
\caption{Histogram and probability density (solid line) of $^{39}$Ar half-life values (see  Table~\ref{tab:Ar39HL})  with addition of 285~years. The vertical dotted line is the mean value and the vertical dashed line is the MFV (they coincide with each other). We have emphasized the specific value of 285 years by applying a darker color to distinguish it from the $^{39}$Ar half-life measurements presented in Table~\ref{tab:Ar39HL}. 
}
\label{fig:Ar39_w285}
\end{figure*}

As another exercise, one can add $T_{1/2} = 285$~years value to the dataset of half-live values presented in Table~\ref{tab:Ar39HL} (one can apply the MVF approach to data shown in Figure~\ref{fig:Ar39_w285}). $T_{1/2}({\text MFV}^{**}) = 268.9^{+3.6}_{-3.1} \, \text{years,}$ with a $1 \sigma$ range of [265.3, 272.0]  corresponding to the 68\% confidence level (a 2$\sigma$ range is [262.1, 275.3] corresponding to the 95\% confidence level). The mean value for the new dataset is 268.9~years (which can be compared with the mean value of 267.7~years for data presented in Table~\ref{tab:Ar39HL}).

\section*{Conclusions} \label{sec:conclusion}

In this work, from the perspective of robust statistics of the observed data, the MFV statistics technique has been applied to explore a detailed statistical analysis for the available dataset of $^{39}$Ar half-life measurements. The MFV estimate for the $^{39}$Ar half-life is $T_{1/2}({\text MFV}) =  268.2^{+3.1}_{-2.9}$ years with uncertainty corresponding to the 68\% confidence level. The uncertainty is  a factor 3 smaller
than that of the most precise re-calculated $^{39}$Ar half-life measurements by Stoenner et al. (1965) (see Equation~\ref{Eq:ar39HL_corr}). Moreover, the consistent results demonstrated the usage, robustness, and tolerance to outliers of MFV statistics relative to mean statistics on an artificial dataset. Following the MFV approach the specific activity of  $^{39}$Ar from underground measurements results in $ SA_{{^{39}\text{Ar}}/\text{Ar}}(\text{MFV}) = 0.966^{+0.010}_{-0.018} \, \, \text{Bq/kg$_{\text{atmAr} } $} $. In addition, the way to estimate $^{39}$Ar half-life from specific atmospheric $^{39}$Ar activity is discussed. However, for that an independent way to determine the fraction of $N_{^{39}\text{Ar}}$ in natural argon (the amount-of-substance fraction) is required. An outlook is provided on how this could be accomplished.

\backmatter

\bmhead{Acknowledgments}

I would like to express my sincere gratitude to Prof. Dr. Werner Aeschbach for the insights provided throughout this research. I am truly grateful for the support and assistance extended by Maria Filimonova. I would also like to extend my appreciation to the management and staff at Canadian Nuclear Laboratories for fostering an enabling environment for this study, with a special mention to Genevieve Hamilton. I am indebted to Helena Rummens for her meticulous editing of the paper. Furthermore, I would like to express my thanks to the anonymous referees for their helpful comments and suggestions, which have greatly contributed to the improvement of this work.

\bibliography{SA}

\end{document}